# On the fluid slip along a solid surface

Jiahao Cheng[1], Jiguang Hao[1*], Yalei Li[1], and J. M. Floryan[2]
[1]School of Aerospace Engineering, Beijing Institute of Technology, Beijing 100081, China
[2]Department of Mechanical and Materials Engineering, Western University, London, Ontario N6A 5B9, Canada

*It is commonly assumed that fluid cannot slip along a solid surface. The experimental evidence generally supports this assumption. We demonstrate that when the change of the relative velocity of a fluid and a solid wall is sufficiently rapid, the slip does occur; the fluid is unable to adjust if acceleration is large enough, and it slips. We use droplet impact on a moving surface to demonstrate and estimate the slip length. We also estimate fluid acceleration, which is required to cause an observable slip.*

Droplet impacts on solid surfaces are ubiquitous in nature and applications, e.g., raining, spraying, printing, coating, cooling, manufacturing, combustion, coffee stain, virus transmission, etc. [1-8]. Investigations in the last three decades were focused on orthogonal impacts but have recently shifted to more practical non-orthogonal impacts and impacts on moving surfaces [9-25].

It is well-known that a tiny air bubble is trapped underneath a droplet impacting a stationary surface [1-5,7,8]. Effects of several parameters, e.g., the liquid property, surface wettability, and ambient pressure, on the bubble dynamics have been investigated [26-38]. The dynamics of a bubble trapped underneath a droplet impacting a moving surface remain to be explored.

Our observations of the trapped bubble document slip occurred during the initial stages of the impact. While the no-slip boundary condition is believed to adequately describe conditions at the solid-fluid contact surface [39,40], there are several exceptions, e.g., rarefied gas [39,41], non-Newtonian fluids [39], Newtonian fluids on superhydrophobic surfaces [39,40,42], and contact line motions [39,43] are known. The slip is modeled by assuming that the fluid velocity tangential to the surface $V_{slip}$ is proportional to wall shear, i.e., $V_{slip} = \lambda \, \partial V/\partial z$ where $\partial V/\partial z$ is the shear rate experienced by the fluid at the wall [39,40,42,44,45], and $\lambda$ is the slip length [45]. $\lambda$ is typically measured using indirect methods, e.g., the pressure drop as a function of flow rate [39,40,46], surface forces apparatus [39,40], atomic force microscopy [39,40,47-49], particles or fluorescence recovery [39,40,42]. $\lambda$'s of up to hundreds of microns have been observed on textured hydrophobic surfaces [39,42,47,50], whereas no direct experimental evidence of a large slip of Newtonian fluids on hydrophilic surfaces has been reported so far [40,49].

A droplet impacting a moving surface has a zero velocity component tangential to the surface. Fluid elements must undergo significant tangential acceleration, infinite in the case of no-slip, which is physically impossible. A finite acceleration implies a slip. The slip time and length can be very small and thus nearly impossible to observe experimentally, especially for methods starting without an initial difference between the tangential velocities of the fluid and the surface [39-50]. In such experiments, the rate of velocity change is insufficient, with the fluid reaching the surface velocity almost instantly.

Consequently, either no slip was observed, or the observed slip was within the measurement error. Placing tracers to observe slip causes errors as tracers move relative to the fluid and may not be at the surface [39]. The impacting droplet on a moving surface provides a way out due to the significant initial velocity difference in tangential velocities and trapping an air bubble between the droplet and the surface whose movement can be tracked using the existing measuring techniques.

Here, we establish a direct method to observe slip and report direct evidence of a large slip occurring when a millimeter-sized Newtonian droplet impacts a moving hydrophilic surface at ambient pressure. An air bubble, whose radius appears not to depend on the surface velocity, is trapped underneath the impacting droplet. Initially, the bubble moves at a lower speed than the surface before eventually reaching the surface speed, i.e., slips along the surface.

The slip length $\lambda$ decreases over time until it reaches zero. The terminal slip distance $l_T$, the maximum distance between the bubble and the impact point, increases with the surface speed and reaches up to 1.39 mm in the range of parameters used in this study. While only the interface between the bubble and the surrounding liquid can be observed, the liquid must also slip along the surface.

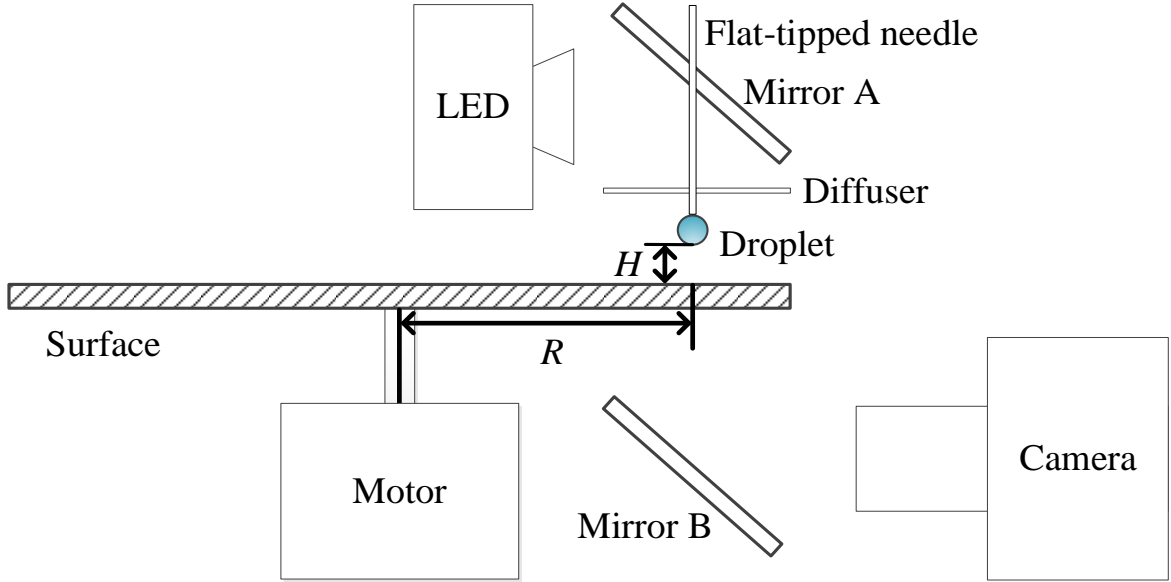

Fig. 1. Schematic diagram of the experimental setup. $R$ is the distance from the impact point to the axis of the rotating surface, and $H$ is the release height of the droplet.

Figure 1 shows a schematic diagram of the experimental setup. The system was designed to permit the independent variation of the droplet speed $V_0$ and the surface speed $V_s$. Ethanol droplets of density $\rho$ = 791 kg/m³, dynamic viscosity $\mu$ = 1.19 mPa s, surface tension $\sigma$ = 22.9 mNm⁻¹ [16,21,24,51], and diameter $D_0$ = 1.77 ± 0.05 mm were generated using a flat-tipped needle driven by a syringe pump and are released from a height $H$ above the surface. As the droplet diameter was close to the capillary length of ethanol [24,33,52], its shape oscillations preceding impact were negligible. Variations of $H$ provided means for varying $V_0$ from 0.76 to 1.26 m/s resulting in the corresponding Weber number $We = \rho D_0 V_0^2/\sigma$ being in the range of 35.1 to 97.4.

The impacted surface was a transparent acrylic disk of radius 11 cm whose rotation was driven by a DC motor with an angular velocity $\omega$ varying between 0 and 17.0 rad/s, resulting in the local surface velocity $V_s = R\omega$ at the impact point at $R$ = 10 cm ranging from 0 to 1.70 m/s, with a precision of ± 0.02 m/s. The equilibrium contact angles of ethanol droplets on the disks were 12.7° ± 1.6 ° determined using ImageJ software.

To observe the trapped bubble, Mirror A was placed above the surface to reflect the LED light through a soft diffuser, and the transparent disk to Mirror B, which reflected it to a Photron Nova S12 high-speed camera equipped with a Navitar 12X long-distance microscope at a pixel resolution down to 5 μm px⁻¹ and at frame rates up to 200,000 fps. This method permits the identification of transient slips like that occurring during acceleration.

**I Bubble formation**

Figure 2 illustrates the effects of $V_s$ on the trapped bubble for $We$ = 54.9. Figure 2 (a) documents the formation of a bubble underneath a droplet impacting a stationary surface, which is consistent with the previous observations [26-38]. Figs. 2 (b)-(e) display images of droplets impacting a moving surface and demonstrate that the air layer trapped underneath the droplet eventually retracts, forming a bubble. The bubble diameter $D_b = (D_1 + D_2)/2$ is independent of $V_s$; here $D_1$ and $D_2$ are the horizontal and vertical bubble's diameters, respectively. The impact point moves faster than the bubble, even for the lowest surface speed of $V_s = 0.38$ m/s. The higher the surface speed, the longer the slip distance $l = (l_1 + l_2)/2$ between the bubble and the impact point. This effect is illustrated in Figs 2 (b) to (e), where $l_1$ and $l_2$ are the distances measured from the left and right edges of the bubble to the impact point, respectively. The existence of $l$ demonstrates that the bubble slips with respect to the moving surface. The bubble does not achieve the surface speed at $T = 200$us, the maximum time used in Fig 2.

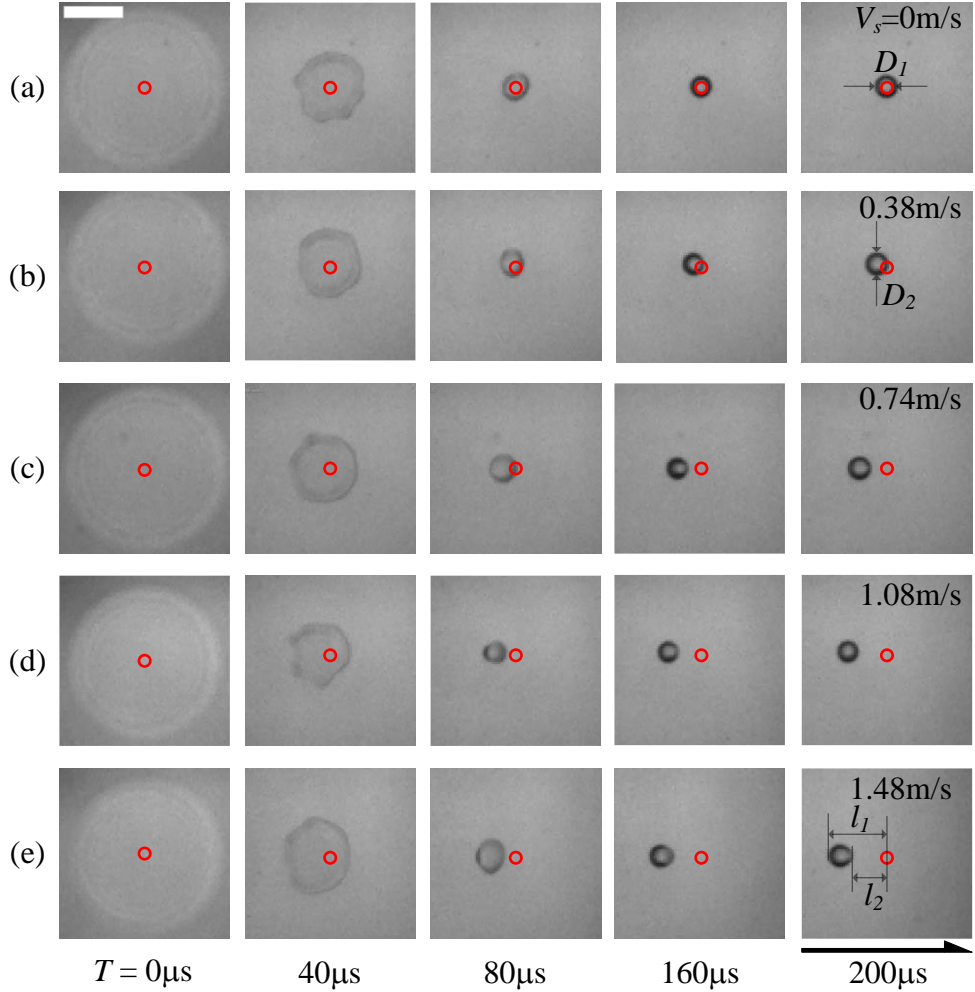

Fig.2 Evolution of the trapped air bubble for various $V_s$'s and $We = 54.9$. (a) $V_s = 0$ m/s, $D_b = 59.3$ μm, $l_{T=200\mu s} = 0$ μm for $T = 200$ μs (Movie S1[53]); (b) $V_s = 0.38$ m/s, $D_b = 60.3$ μm, $l_{T=200\mu s} = 26.9$ μm; (c) $V_s = 0.74$ m/s, $D_b = 59.4$ μm, $l_{T=200\mu s} = 63.8$ μm; (d) $V_s = 1.08$ m/s, $D_b = 58.3$μm, $l_{T=200\mu s} = 94.2$ μm; (e) $V_s = 1.48$ m/s, $D_b = 60.4$ μm, $l_{T=200\mu s} = 110.1$μm (Movie S2[53]). Red circles identify the locations of the impact points. The scale bar is 140 μm. The arrow at the bottom indicates the direction of surface motion.

Figure 3 illustrates the evolution of the trapped air film for various $We$'s and $V_s = 1.48$ m/s. The bubble diameter $D_b$ decreases from 62.0 μm to 58.0 μm as $We$ increase from 35.1 to 76.3, consistent with the previous results dealing with droplets impacting a stationary surface. When the air film retracts into a bubble starting from the moment of contact, the time $T_b$ required for the test variable $|1 - D_1/D_2|$ to reach 0.05 decreases from 168 μs to 115 μs for the same range of $We$, indicating that retraction is faster when a smaller amount of air is trapped. The observed slip distance $l$ weakly depends on $We$, which further supports a large slip (see Fig.2).

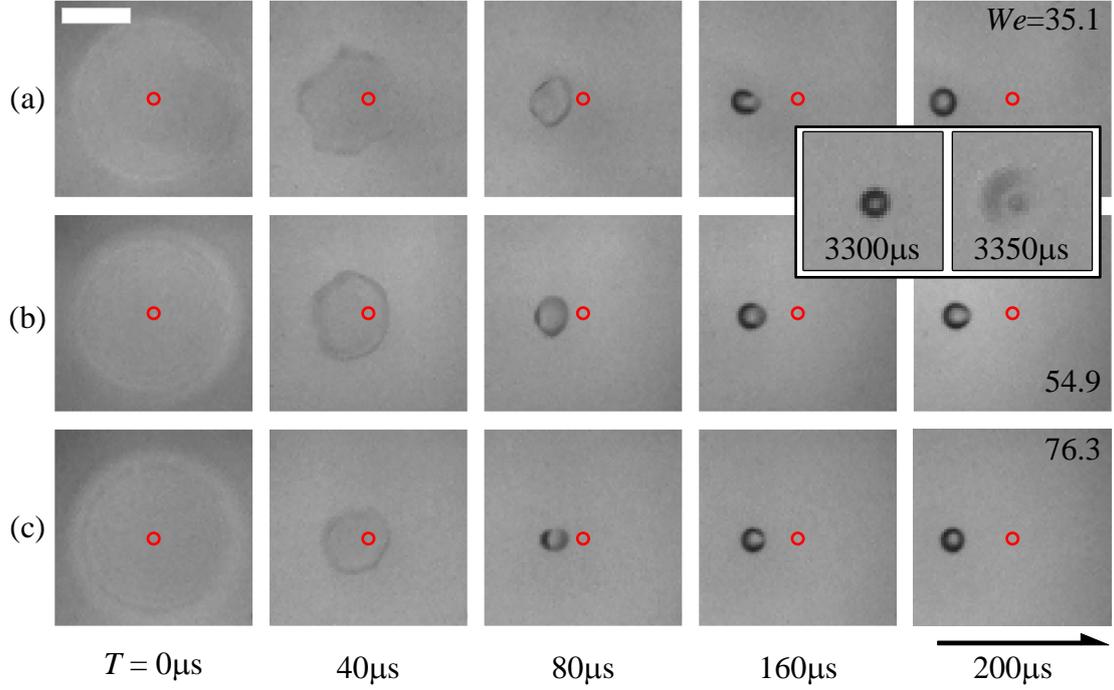

Fig.3 Evolution of the trapped air bubble for various *We*'s and $V_s$ = 1.48 m/s. (a) $We$ = 35.1, $D_b$ = 62.0 μm, $l_{T=200μs}$ = 140.2 μm; the inset demonstrates bubble existence at $T$ = 3300 μs and its breakup at $T$ = 3350 μs; (b) $We$ = 54.9, $D_b$ = 60.4 μm, $l_{T=200μs}$ = 114.9 μm; (c) $We$ = 76.3, $D_b$ = 58.0 μm, $l_{T=200μs}$ = 121.8 μm. The scale bar is 140 μm.

Figure 4 shows variations of time $T_b$ required for the bubble formation (a) and the bubble diameter $D_b$ (b) as functions of $V_s$ and *We*. $T_b$ and $D_b$ are the average values of three experiments under identical conditions, respectively. $T_b$ and $D_b$ decrease as an increase of *We*, consistent with the previous study [54]. Fig. 4 (b) shows that $D_b$ is independent of $V_s$.

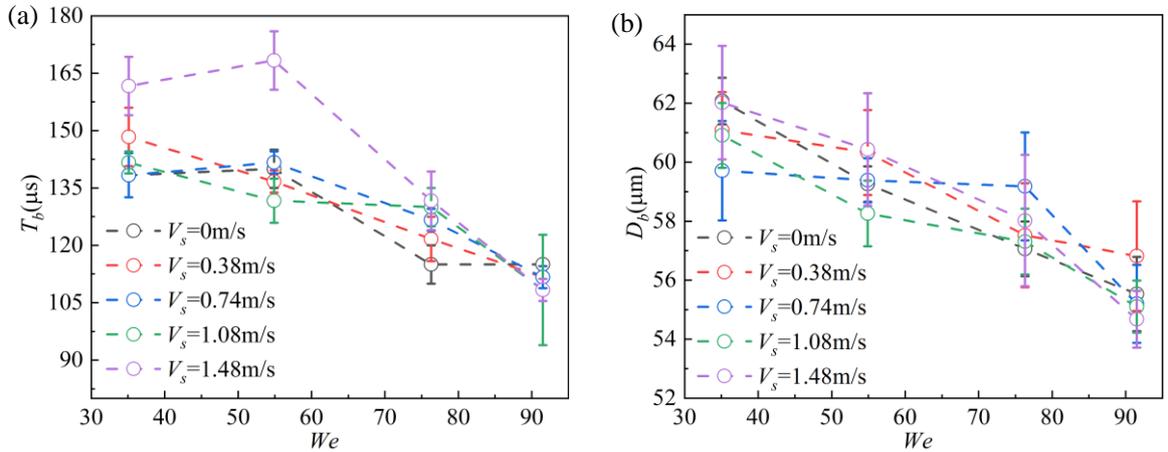

Fig.4 Variations of the bubble formation instant $T_b$ (a) and the bubble diameter $D_b$ (b) as functions of $V_s$ and *We*. Error bars indicate the standard deviations.

**II Bubble acceleration**

Figure 5 illustrates the bubble acceleration process for various $V_s$'s at $We$ = 97.4 well above $We$ = 55, as otherwise, the bubble would detach from the surface, rise to the interface, and collapse, as demonstrated in the inset in Fig.3 (a). The position of the impact point on the moving surface was determined by multiplying $V_s$ and $T$ and is marked using a red dot. A red dashed line marks its relative position with respect to the bubble. The slip distance $l$ increases over time, indicating that the surface moves faster than the bubble. We define the terminal slip distance $l_T$ as the average of ten $l$'s determined after the time $T_R$ required for the bubble to reach speed $V_s$ with an error of 5%, as shown in Fig. 6 (c). Results displayed in Figs.5(a) to (c) and Fig.6(a) and (c) illustrate how $l_T$ increases as an increase of $V_s$.

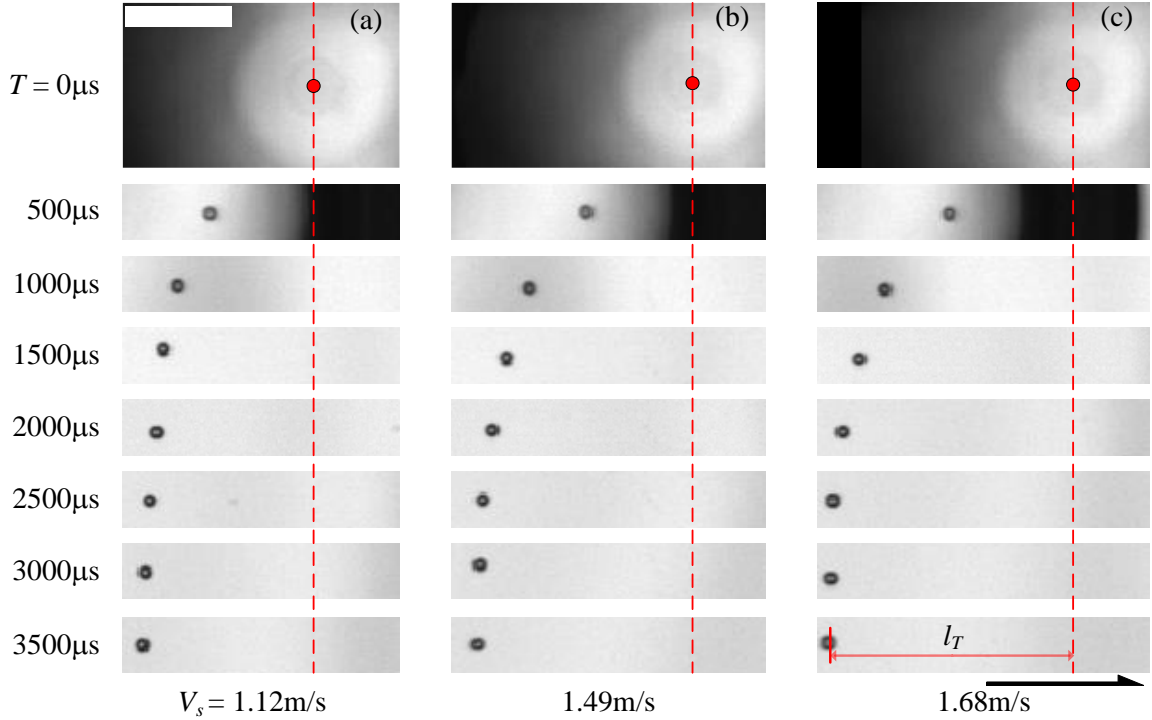

Fig.5 Evolution of the trapped bubble for $We = 97.4$ and various $V_s$'s. Red dashed lines and points identify the positions of the impact points on the moving surfaces. (a) $V_s = 1.12$ m/s, the terminal slip distance $l_T = 439.0$ μm (Movie S3[53]); (b) $V_s = 1.49$ m/s, $l_T = 710.6$ μm; (c) $V_s = 1.68$ m/s, $l_T = 794.3$ μm (Movie S4[53]). The scale bar is 370 μm.

Figure 6 shows $V_b$ (a), $T_R$ (b), $l$ (c), and $l_T$ (d) of the bubble. The bubble displacement $S$ (see the inset in Fig. 6(a)) is determined from the bubble positions extracted from the high-speed images; the time derivative of the displacements determines the bubble speed $V_b$ (see Fig.6 (a)). Figure 6 (a) shows that $V_b$ is smaller than $V_s$ at the initial impact stage and increases until reaching $V_s$ at $T_R$; subsequently, the bubble moves at $V_s$. These results demonstrate slip occurring during acceleration until the bubble reaches $V_s$. Figure 6 (b) indicates that $T_R$ increases as $V_s$ increases and weakly depends on $We$. Figure 6 (c) shows that the slip distance $l$ increases over time and reaches the final value of $l_T$ after $T_R$. Variations of $l_T$ as a function of $We$ and $V_s$ shown in Fig. 6 (d) further indicate that $l_T$ increases as an increase of $V_s$, whereas $We$ weakly influence $l_T$. $T_R$ and $l_T$ are the averages measured during three experiments under identical conditions.

Experiments [39-50] showed no observable slip in fluid flow along a hydrophilic surface when one of the initially identical tangential velocities was altered. A slip must occur for a transient flow where a difference between tangential velocities exists initially, as no slip implies infinite acceleration, which is physically impossible. The fluid is accelerated to the surface velocity within a specific time interval. Our results show that this interval is of an order of milliseconds. During that time, a large slip occurs until the fluid reaches $V_s$. Afterward, our results matched the results from the literature, and no slip is observed. At the beginning of the impact, $\lambda = \infty$ as $V_{slip} = V_s - V_b = V_s$ as $V_b = 0$; at the end of the impact $\lambda = 0$.

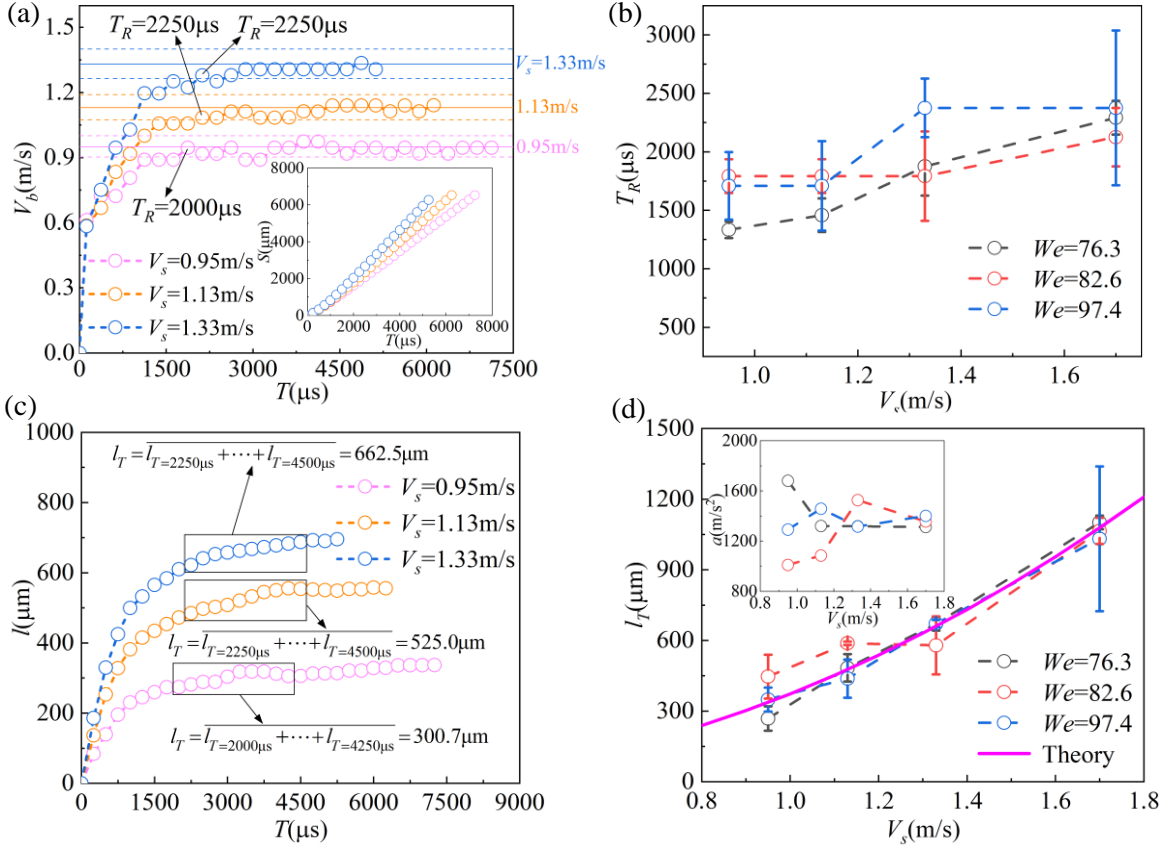

Fig.6 (a) Variations of the bubble speed $V_b$ as a function of $T$ and $V_s$ for $We = 97.4$. The blue, orange, and magenta solid lines represent $V_S = 1.33$, 1.13, and 0.95 m/s, respectively, while the dashed lines of the same colors represent ±5% of the $V_s$. The inset illustrates the variations of the bubble displacement $S$ as a function of $T$ and $V_s$. (b) Variations of $T_R$ as functions of $V_s$ and $We$. (c) Variations of the slip distance $l$ as a function of $T$ and $V_s$ for $We = 97.4$. (d) Variations of $l_T$ as a function of $V_s$ and $We$. The inset illustrates the experimentally determined acceleration $a$.

Assuming that the bubble acceleration $a$ is constant during the acceleration process, $a$ determined using $a = V_s^2/(2l_T)$ and the experimentally-determined $l_T$ is in the range of 1010 to 1681 m/s$^2$ (see the inset in Fig. 6(d)); the average $a$ is 1340 m/s$^2$. Using $a = 1340$ m/s$^2$, $l_T$ can be theoretically determined as $l_T = V_s^2/(2a)$ with results illustrated using the solid magenta line in Fig. 6(d). The predicted $l_T$ agrees with the experimental data reasonably well, indicating that the acceleration process can be reproduced using a constant acceleration, thus providing a physical explanation of the increase of $l_T$ with $V_s$ and a simple tool for predicting $l_T$ and the time when the slip length changes from infinite to zero. The 1340 m/s$^2$ is an average value; it is larger initially but smaller at the later stage of the slip and is more than two orders of magnitude larger than the gravitational acceleration.

In conclusion, we experimentally investigated slip occurring during droplet impact on a moving hydrophilic surface. The relevant information was extracted by observing the movement of the air bubble trapped underneath the impacting droplet. The bubble and the surrounding liquid slip on the surface, with the slip length being initially infinite and then decreasing to zero, demonstrating that slip, even a perfect slip, exists during fluid acceleration. In contrast, no slip is observed after the acceleration decreases below a certain level. The terminal slip distance $l_T$ was up to 1.39 mm in the parameters' range used in this study; it increases with an increase of $V_s$ and depends weakly on *We*. Assuming that the bubble acceleration is constant, the measurements show an average acceleration of 1340m/s$^2$, which explains the increase of $l_T$ with $V_s$ and can be used to predict $l_T$ with reasonable accuracy.

All previous experimental methods studying slip started with fluid and solid surfaces moving with the same tangential velocity and then accelerating one of them but with an acceleration much smaller than 1340m/s$^2$. The fluid can adjust to such acceleration leading either to no observable slip or the observed slip being within the measurement error. To the best of our knowledge, the experimental setup used in our experiment is the only one that produces acceleration large enough to cause an observable fluid slip and quantify this slip.

**ACKNOWLEDGMENTS**

This study was financially supported by the National Natural Science Foundation of China under Grant No. 12072032, the National Key R&D Program of China under Grant No. 2018YFF0300804, and 111 Project under Grant No. B16003.

*hjgizq@bit.edu.cn